TITLE:  Empirical Photometric Control of Mars Context Camera Images

RUNNING HEAD:  Empirical Photometric Control of CTX

Stuart J. Robbins*[,a]     ORCID: 0000-0002-8585-2549

Michelle R. Kirchoff[a] ORCID: 0000-0002-3367-2730

Rachael H. Hoover[a]     ORCID: 0000-0003-0926-7513

*stuart@boulder.swri.edu

*Corresponding Author

[a]Southwest Research Institute, 1050 Walnut St., Suite 300, Boulder, CO 80302, United States

Total Pages – 23; Total Tables – 0; Total Figures – 5

Key Points:

- Provide the first broad-scale application of empirical photometric control to Context Camera data.
- Resulting mosaics are nearly seamless when images are relatively consistent.
- Accuracy of control depends upon accuracy of a reference source.

Plain Language Summary (200-word limit):

Creating a mosaic of images of planetary surfaces is relatively straightforward: Geometric data of the camera and spacecraft tell us where and how images should be placed relative to each other. Something that is more difficult is matching brightness variations across the images to create a mosaic that appears seamless. Different mathematical mechanisms exist to try to adjust brightness and contrast of overlapping images to match, but this is not always possible. For example, if two overlapping images were taken when the sun was in different locations in the sky, then it will be impossible to get the shadows to match. If clouds or a dust storm dims an image and reduces contrast, it will be hard to match to another that was taken when the atmosphere was clear. In this work, we present a more advanced method to create the appearance of a seamless mosaic. We achieve this by not matching images to each other, but to a coarser, lower resolution reference that already has even illumination. We tailored our application to the Context Camera that has been returning ≈6 meters per pixel images of Mars since 2006, and we demonstrate it across different terrains of Mars.

Abstract (250-word limit):

The *Mars Reconnaissance Orbiter* (*MRO*) spacecraft has been in orbit around Mars since March 2006.  The Context Camera (CTX) on *MRO* has returned over 100,000 images of the planet at approximately 5–6 meters per pixel, providing nearly global coverage.  During that time, Mars has gone through nearly 7 of its own years, changing solar distance from 1.38 to 1.67 AU and the corresponding solar flux by 45% due to its orbital eccentricity.  Seasonal effects and transient phenomena affect atmospheric transparency.  Combined with an aging detector, CTX images are difficult to mosaic seamlessly, for all of these changes prevent equalizing images to create a visually smoothly illuminated product.  We have developed a method, based on previous work by other researchers for other datasets, to mitigate almost all photometric variations between images in order to create the appearance of an evenly illuminated, *practically* seamless mosaic.  We describe how the process works, which uses a reference source to tie brightness values, and demonstrate its effects across Mars' surface.  While the workflow developed for this product is applicable to other planetary bodies, it requires a reference source, which may not yet exist.

## 1. Introduction

Mars was first visited by *Mariner 4* in 1965, and since then, numerous cameras have flown on flyby, orbiter, and rover missions to the red planet. One of the primary products of any camera system on such craft is the photographs from which scientists can interpret the surface. Interpretations are greatly improved when images provide consistent lighting and illumination. For Mars, the atmospheric processes and elliptical orbit present a rare problem for creating a smoothly illuminated, imaging artifact-free surface.

Six missions have produced high-resolution (better than 1 km/pix), global or nearly global image datasets of Mars' surface: *Mariner 9*'s Visual Imaging System, *Viking*'s Viking Orbiter Camera, *Mars Global Surveyor*'s Mars Orbiter Wide-Angle Camera (MOC-WA), *Mars Odyssey*'s Thermal Emission Imaging System (THEMIS), *Mars Express*'s High-Resolution Stereo Camera (HRSC), and *Mars Reconnaissance Orbiter*'s Context Camera (CTX).

No global mosaic was created from *Mariner 9*, which was quickly superseded by *Viking 1* and *2*. The original *Viking* mosaics were utilized for decades. An improved version, *Viking* MDIM 2.1 (Archinal *et al.*, 2003), put the 232 mpp mosaic into the modern cartographic control system and made concerted efforts to create a photometrically useful (roughly seamless) product, though seams and mismatches remain.

For modern spacecraft, a nearly global mosaic was produced from MOC-WA at 256 ppd from its initial mapping orbit (Caplinger and Malin, 2001). It did not initially include south of ≈60°S because the mapping phase was during Mars' north polar summer, though south of 65°S was later produced as a supplement. There has yet to be a global mosaic made from HRSC data, though various reports have suggested it is forthcoming, including one that formed the basis for this work (Michael *et al.*, 2016). A controlled THEMIS mosaic has just been released within 65° of the equator as a relatively seamless, fully controlled product (Ferguson and Weller, 2018). Uncontrolled THEMIS is available globally.

The lack of a visible-light, global, high-resolution (better than ~20 mpp) mosaic formed the basis for our related cartographic control work, with the primary goal to produce a global CTX mosaic that is both controlled in a relative sense (image-to-image) and absolute sense (image-to-reference, the Mars Orbiter Laser Altimeter (MOLA) datum). Our cartographic work produces a mosaic that is seamless with respect to features being continuous across image boundaries (Robbins *et al.*, 2019, in rev.); however, jarring brightness differences at image edges due to effects described above can make it appear incorrect and more difficult to use. That formed the motivation for this work: To develop a method to produce the appearance of a photometrically stable, approximately calibrated product with approximately accurate relative reflectivity.

We developed a workflow to accomplish that goal of a photometrically stable and consistent product. This workflow produces a complementary – but very different in practice – product to the "seamless" CTX mosaic released by Dickson *et al.* (2018). Their efforts focused on judiciously choosing what pixels from each image to use in a mosaic based on clipping borders at minimum contrast features. For example, if a crater appears near an image seam but exists entirely on one image and not the other, their methods will preserve the pixels from one image with the entire crater, even if the other image overall might otherwise be placed on top of the mosaic. Combined with equalization, their product is a good approximation of what the human eye might see.

Our product differs by including a different photometric calibration step and applying that to every image before mosaicking, which we describe in section 2. Our process is heavily dependent on a reference brightness map, so we discuss plausible sources for calibration in section 3. Section 4 shows examples of the calibration, and section 5 presents a brief discussion.

## 2. Methods

### 2.1. Initial Image Processing

Figure 1 illustrates our workflow. This section describes the Level 1 processing. The first several steps are standard for processing CTX data in the United States' Geologic Survey's (USGS's) *Integrated Software for Imagers and Spectrometers v3* (*ISIS3*), based on online

documentation:

- MROCTX2ISIS:  Imports Planetary Data System (PDS) CTX images into *ISIS3*.

- SPICEINIT:  Attaches SPICE data to the image.

- CTXCAL:  Applies *ISIS3*'s geometric and radiometric calibrations to the image.

- CTXEVENODD:  Corrects for even-odd line differences inherent to CTX.

Figure 2A shows a mosaic of Tharsis Tholus after the above steps.  An additional, rarely seen photometric calibration step is an empirical flat field to CTX images.  As shown in the top row of Fig. 2, most noticeable in panels B-C, the CTX detector has a well-known "smile" (or "frown") where pixels at the edges of the detector are less sensitive than those in the middle, resulting in darkening across the linescan camera.  The effect is not removed with CTXCAL, and it has changed during the lifetime of the camera.  We apply a technique common in astronomical observations of an empirical flat field:

- MAKEFLAT:  Reads $N$ rows of every image with the same $M$ pixel width, and median-combines them to make a $1 \times M$-pixel empirical flat.

- RATIO:  Takes the ratio of this flat relative to each image that is $M$ pixels wide, along the entire image track, normalizing across-track brightness variations.

The result is in the bottom row of Fig. 2 (the different columns represent different mosaic mechanisms, described in the next sections).  The process must be repeated for each different $M$ image widths, and we used $N = 5000$ pixels (the detector's RAM can store images up to ~50,000 pixels long, or ~100,000 pixels if the across-track summing was set to 2).  We found $N = 5000$ pixels a good compromise between computer time and correction quality.  We ran this step in batches in temporal space to account for sensitivity changes as the equipment aged, mostly based on the first three characters of the image file name (the first letter corresponds to Mars Year, and the two numbers correspond to Earth month in that Mars Year).

There were still some cases where this failed to give a visually correct result: If areas of significant brightness differences occurred in the first 5000 pixels, vertical streaks of improper correction would occur in the rest of the image (not shown); this could affect an unrelated image that had the misfortune of also having *M* columns. Images were visually examined and the few that had this effect were re-processed with different parameters.

## 2.2. Methods that Did Not Work with CTX: Equalization and Averaging

We briefly describe two simpler methods than our eventual solution that did not work with CTX data in order to provide context for why we went with a more involved technique.

First, we attempted image equalization (Fig. 2B-C, E-F), which tries to match the overall brightness of images by shifting (brightness increase or decrease) and scaling (contrast increase or decrease). However, CTX images are not just mismatched in brightness via a DN offset and/or contrast scale/stretch, but also in the image scene itself, from one image to the next. For example, with swaths 30 km wide and up to 300 km long, transient and localized atmospheric phenomena change the brightness over parts of an image but not others, such that image edges might match over some parts but not over others. As another example, images taken close to the terminator (but not actually over it) can have an overall gradient along track – or at an angle – that can be almost impossible to model and remove. Finally, without anchors, equalization will tend to introduce long-wavelength brightness undulations across the final mosaic. Anchoring with "held" images is itself arbitrary and requires selecting several anchors and scaling them to an *a priori* "correct" brightness. While *ISIS3*'s LINEEQ could be used to mitigate some along-track difference, it is difficult to use correctly, for it can easily remove real albedo differences that should be preserved.

Second, we examined equalization with averaging (Fig. 2C,F). The idea of averaging is, if multiple images exist for a certain location on a planetary surface, and they have been normalized to a similar brightness via equalization or careful photometric correction from first principles, then an average of the pixel values at that location could mitigate transient effects and produce a visually even product. Unfortunately, there are issues with an average mosaic from CTX data. First, except in very limited locations on Mars, there do not exist enough CTX images to broadly apply this technique. Over much of the surface, only one image exists, and in some areas there remain gaps in coverage with usable images (at the time of this writing, with PDS release 49, there is ≈98.1% global CTX coverage with usable images, where a "usable" image is one where surface features are visible and there is good signal-to-noise). Second, while CTX images were taken at approximately the same time of day, all seasons are present in the catalog. Mars has an obliquity of ≈25°, meaning that over one year, shadows shift by up to 50°, making images inconsistently illuminated for an average. The third issue is that, while most CTX images were taken close to nadir pointing, slews were done by up to ~40° in either direction. Even with topographic correction from MOLA, feature mismatches remain with these slews such that an average produces noticeable ghosting. Fourth, Mars' appearance does change, and in non-periodic ways, such that averaging can still produce a seamless scene (see lengthy discussion in Robbins, in rev., and references therein), but mask important, transient features that may be of interest at CTX scales.

## 2.3. Using a Reference for Empirical Calibration

The method we found worked best, and thus further developed specifically for CTX, is based on methods detailed by Michael *et al.* (2016) for HRSC calibration. The principle as applied to Mars mosaics is used at least as far back as MOC-WA by Caplinger and Malin (2001). The basic theory is straightforward: Every input pixel from a CTX image is compared against a reference, and it is scaled to that reference to match it. The workflow we developed for this is shown in Fig. 1. Note that some software will treat the parameters differently (*e.g.*, σ in a Gaussian blur), such that other implementations may vary. We used Python, its SciPy package, and GDAL's python implementation for the bulk of this process, in combination with *ISIS3*.

For the initial steps, which are in *ISIS3*, we create map-projected versions of each CTX image at 1000, 50, and 6 mpp (with 6 being the scale at which we produce mosaics). Each map-projected CTX image will have null pixels at the edges because the images will no longer be perfect rectangles. Additionally, we map-project (into the same projection system) the reference mosaic at 1000 and 50 mpp. We then extract a latitude × longitude range from the 1000 mpp reference for each CTX image, effectively creating what the CTX footprint "should" look like from the larger reference.

In Python, the 1000 mpp CTX and reference images are ratioed. Null pixels at the edges are interpolated via SciPy's nearest neighbor interpolation – just DN values that are reasonably similar to the content of the image are needed (where "reasonably similar" is a qualification that must be made by the individual researcher, though we found no differences in the final product when DN values were within a factor of ~1.5). However, this ratio alone would remove most real brightness data. Instead, we smooth the ratio with a Gaussian blur with σ=5 (5 km at this scale). This blur is why the null pixels at image edges must be filled: if unfilled, the null pixels will intrude into the image pixels when blurred, and if filled with 0-value data, then the edges will be darkened. The blurred ratio is up-sampled by 20×, which is used to equalize the 50 mpp CTX image through simple division. The blurred ratio image provides three important qualities: (1) The long-wavelength anchor that is missing from standard equalization; (2) it is short enough in wavelength that it will correct for most issues across a single CTX image; and (3) it is long enough in wavelength that it allows short-wavelength, real albedo structure to remain.

Images are converted back to *ISIS3*, and that software is used to build a 50 mpp mosaic. Because there will almost certainly be gaps in coverage, the 50 mpp mosaic is placed on top of the 50 mpp reference product such that where the CTX data exists, those pixels will remain, but where there are gaps, the reference shows through (if no gores were present, this step could be skipped). The result is degraded again with a Gaussian blur of σ=35 pix (1.75 km) in Python (Python libraries are more efficient than the *ISIS3* Gaussian blur). This result is treated as a new brightness reference, 50 mpp rectangles are extracted from it to cover each CTX image, and ratios between them and the original 50 mpp CTX are constructed. These ratios are blurred by 50 pixels (2.5 km), up-sampled, and applied to the 6 mpp images. These 6 mpp images are then mosaicked and considered to be the final product. The purposes of the intermediate mosaic and second calibration are two-fold: First, the low-resolution reference mosaic calibration is not perfect due to its own photometric issues, different wavelength of light sensitivity, and resolution (see next section); and second, image edges and some mismatches remain from the first calibration that are mostly removed in the second, though some edges are still visible.

It is possible that an improved mosaic could be constructed with a third pass, but this approaches the point of diminishing returns. Map-projected, 6mpp CTX images can be >5 GB, and so it is not trivial, even on a modern computer with dozens of CPUs and NVMe storage, to process hundreds (or tens of thousands) of CTX images >2 times. Several of the parameter choices we made are based on subjective appearance and, when modified slightly, tend to produce similar results. Specifically, we were looking for parameters that appeared to preserve short wavelength brightness variations (at a scale much smaller than an individual CTX image) while eliminating long wavelength variations (at the scale of a single CTX image). Admittedly, this is vague, but this is a subjective process since a true photometric reference does not exist at CTX scales.

Our work varies from Michael *et al.*'s implementation in several ways, primarily due to the different requirements between CTX and HRSC. One is the choice of scale, where the intermediate mosaic used as their second albedo reference product is 400 mpp, but ours is 50 mpp. In addition, they down-sample the original reference and comparison to approximately 3 pixels ("cells" in their description) wide, while we maintain the projection at 1000 mpp such that CTX images are ~30 pixels wide; their second stage correction uses 9 pixels, while we use ~600. We have found this produces a better, smoother result for CTX. Another difference is that they performed all calculations using 16-bit integers for computational speed, while we maintain all calculations in 32-bit floating for, again, a smoother product. The final primary difference is the reference source, which is the subject of the next section.

## 3. Reference Brightness Map

Figure 3 shows the results of using different reference maps. There are only two true bolometric Bond albedo map of Mars: One was produced from *Mars Global Surveyor*'s Thermal Emission Spectrometer (TES; Christensen *et al.*, 1992, 2001), and the other was produced from *Mars Express*'s Observatoire pour la Minéralogie, l'Eau, les Glaces et l'Activité (OMEGA; Bibring *et al.*, 2004; Vincendon *et al.*, 2015; Audouard *et al.*, 2017). The visible/near-IR bands, 300–2900 nm, were used to generate the TES map. Unfortunately, it is not controlled to the current Mars datum; is extremely low-resolution (8 ppd or 7.5 km/pix, which is >1000× coarser than CTX); and even at that low resolution, it has significant gaps, interpolation artifacts, and does not cover either pole. Its much broader spectral response is also not ideal for calibrating a camera sensitive to 500–700 nm. We found it could not preserve even large-scale (~½ image-width) CTX reflectivity and introduced significant brightness artifacts that manifest as splotches the size of a TES pixel. While OMEGA is higher resolution (60 ppd or 1.0 km/pix, which is ~200× coarser than CTX), it still has significant artifacts in the form of mismatches at image edges, which manifest as patches of much darker or lighter regions. Our own investigation also shows that it might have overall gradients present that would affect our product. Therefore, we found neither of these products reasonable for producing the type of smooth CTX mosaic desired.

We did not investigate THEMIS due to its sensitivity centered at 12,570 nm, despite it now being fully controlled for most of Mars (Christensen *et al.*, 1991; 2001; Edwards *et al.*, 2011; Fergason *et al.*, 2019). *Viking* is also controlled (Archinal *et al.*, 2003), is visible-light, and is *generally* photometrically consistent (*e.g.*, Carr *et al.*, 1976) though issues remain (Soderblom, 1978). However, while it is significantly better for this type of effort with its higher spatial resolution, it still has photometric issues, and it is not an albedo map: Highlights and shadows in craters are prevalent, and because data were not taken at the same time of day as CTX (Fig. 4), CTX would be "corrected" to have a bright highlight when there should be a shadow (Fig. 3E,F).

We instead use MOC-WA, a relatively little-used dataset, but close to ideal for this calibration effort: The public MOC-WA mosaic is photometrically stable; was taken at 575–625 nm (exactly centered in CTX's range); and, importantly, images were taken at nearly the same time of day as CTX, averaging just 1.2 hours earlier (Fig. 4). MOC-WA images are close to local noon, making shadows shorter and highlights less extreme, bringing it nearer to a true normal reflectivity map. Although it does have some issues with seam mismatches, is not registered to MOLA, and has a gap 60–65°S, it still produces a result superior to TES or *Viking* (Fig. 3G,H). It is also relatively high resolution, 256 ppd (232 mpp at the equator; ~40 CTX pixels wide). For those reasons, we suggest that it is the best current option for performing this empirical calibration for CTX. However, to mitigate the 60–65°S gap issue, and extreme shadows in the <65° mosaic, we separately produced our own version of a MOC-WA red normal reflectivity map, using the limb-to-limb imaging (9 ppd, 6.6 km/pix), for the narrow $L_s$ = 270° (southern solstice), and describe that product in separate work (Robbins, in rev.). The much lower resolution means that the effective "cell" number to calibrate the 1000 mpp CTX images is ≈5 pixels across, close to Michael *et al.* (2016).

## 4. Results *via* Examples

In Figure 3, we showed the result of our process in a small area made of 42 images. We also demonstrate our process for four different terrain types with many more individual images in Figure 5: Volcanic (Fig. 5B), plains (Fig. 5B,D), tectonic (Fig. 5D), and south polar (Fig. 5F). Regions of Mars that vary relatively little in reflectivity produce excellent results even with gores (Fig. 5B), and it does not create extra contrast where none exists (*e.g.*, the washed out images in the upper-right are preserved as low contrast). Even areas that have some secular variation produce a visually even product (Fig. 5D), though some small areas of mismatch do remain.

Areas with large, real variations in reflectivity do not produce a perfect product (Fig. 5F), though it is significantly improved over a raw mosaic (Fig. 5E) or one with equalization applied (not shown). In this south polar example, seasonal frost covers some features in some images, but not the same features in adjacent images. Large contrast between the frost and regolith are preserved through our technique, which visually clashes with adjacent images where the frost has sublimated so the terrain has less contrast.

## 5. Summary and Discussion

In this work, we presented a technique tailored to produce an empirical photometric correction for CTX data. It is based on past efforts designed for other instruments, and it shows improvement over what is commonly applied in the field. The quality of photometric control is dependent on the quality of the reference source used, and for CTX we argue that a MOC-WA product is currently the best reference. However, it does have issues that we are looking to potentially mitigate through other work. For example, the photometric reference used for the south pole was made by correcting limb-to-limb MOC-WA and averaging hundreds of images together, producing a much smoother and consistent reference, though secular variations in Mars' reflectivity prevent broad applications. Producing additional maps for other $L_s$ windows and using the closest $L_s$ window to each CTX image may produce a better calibrated product. In future work, control to the MOLA datum would make MOC-WA more consistent with our CTX effort. Exploration of whether contemporaneous near-global imaging would produce a more consistent product (such as from MARCI – the Mars Color Imager; Malin *et al*., 2001; Bell *et al*., 2009), given Mars' temporal changes, is also an avenue of future exploration. Meanwhile, we have applied this correction to version 1 of our south polar CTX mosaic (Robbins *et al*., in rev.). While this workflow and reference base are tailored specifically for CTX, it is certain that, as a general workflow, they can be adapted to other datasets.

Acknowledgements: The original Mars Reconnaissance Orbiter Context Camera imagery used in this work are available in NASA's Cartography and Imaging Sciences Node (IMG) of the Planetary Data System (PDS): https://pds-imaging.jpl.nasa.gov/volumes/mro.html . The TES map used for photometric control is available from the Mars Global Surveyor Thermal Emission Spectrometer website: http://tes.asu.edu . The Mars Orbiter Camera (narrow-angle) mosaics used is available from Malin Space Science Systems: https://www.msss.com/mgcwg/mgm/ . The Mars Orbiter Camera (wide-angle) mosaic used for south polar control is available from NASA's Cartography and Imaging Sciences Node (IMG) of the Planetary Data System (PDS) "Annex" housed at the United States Geologic Survey: [link to be provided upon acceptance of that work in JGR] (Robbins, in rev.). The Viking mosaic used is available from NASA's IMG PDS Annex: https://astrogeology.usgs.gov/search/map/Mars/Viking/MDIM21/Mars_Viking_MDIM21_Mosaic_global_232m . The OMEGA mosaic used is available from the Planetary SUrface Portal (PSUP): http://psup.ias.u-psud.fr/sitools/client-user/index.html?project=PLISonMars (Audouard *et al.*, 2017). The authors thank T.M. Hare for useful discussions about several aspects of the control and photometric correction process. This work was funded through internal awards to the authors by Southwest Research Institute.

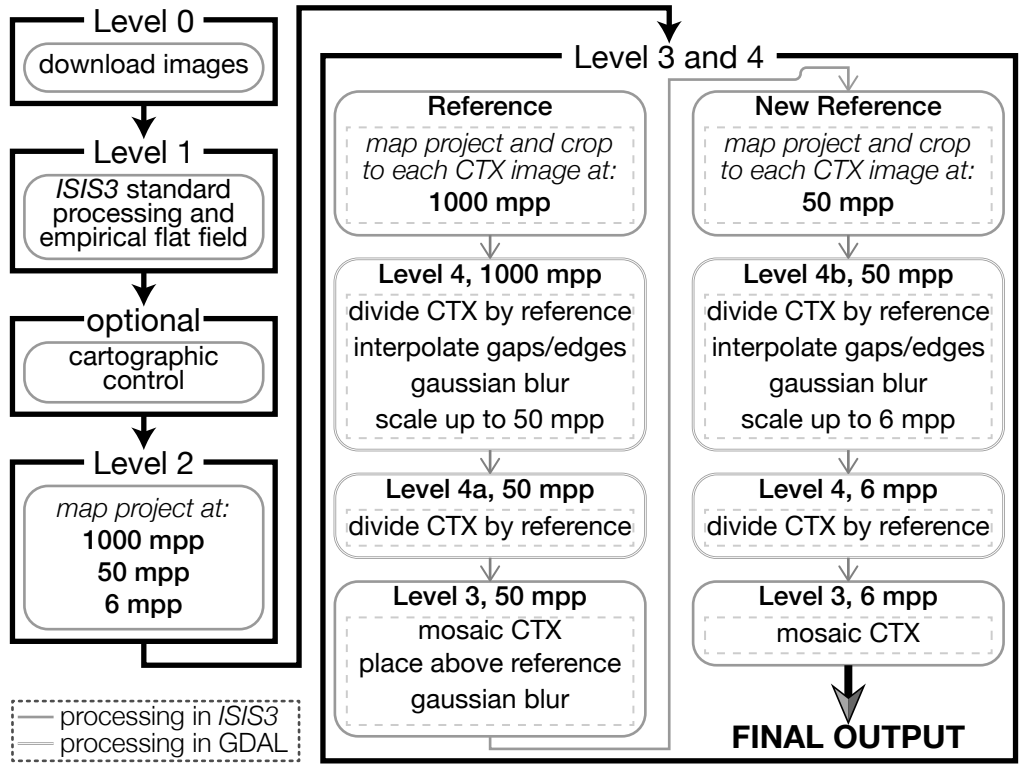

Figure 1. Flowchart illustrating the workflow to produce empirically photometrically controlled CTX mosaics. This work does not require CTX images to be cartographically controlled. Standard processing terms are used: Level 0 are raw images, Level 1 are calibrated images, Level 2 are map-projected, Level 3 are mosaicked, and Level 4 are photometrically controlled.

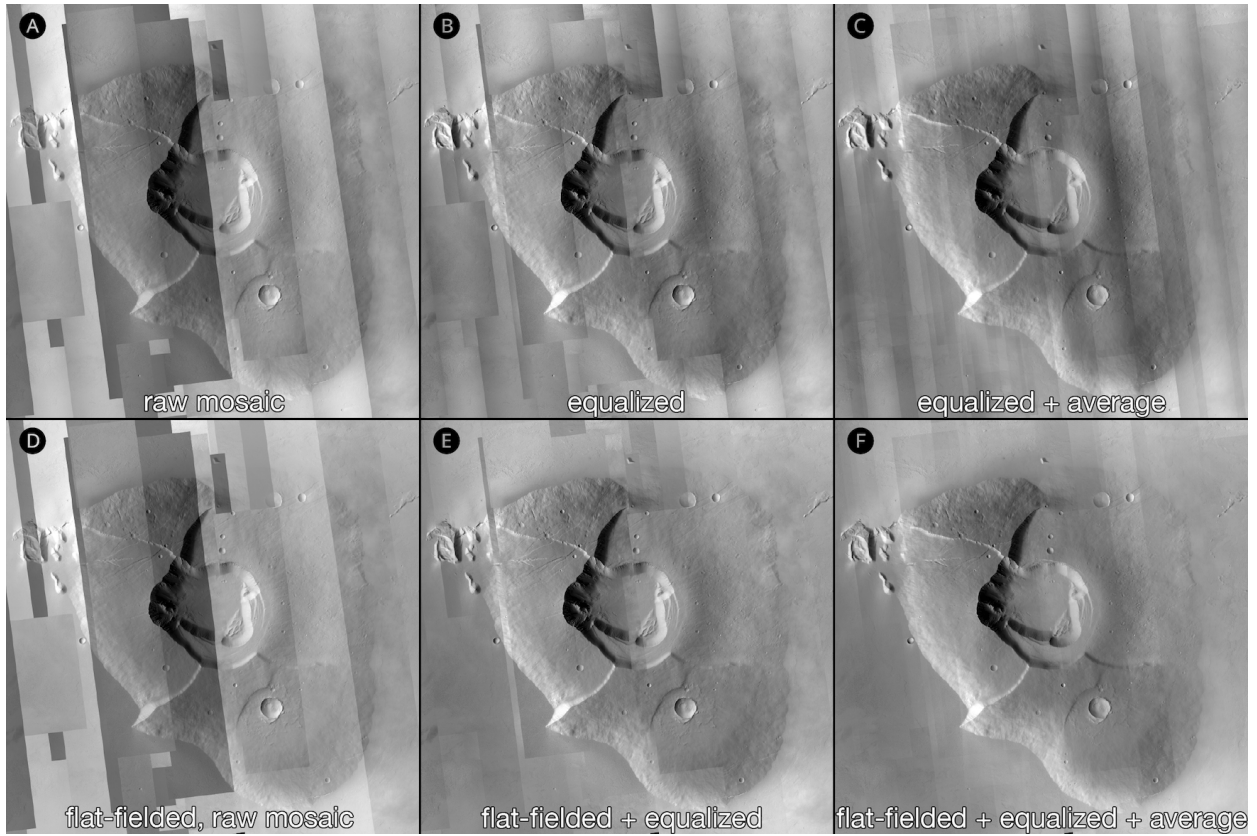

Figure 2. Example of initial calibration steps and equalization effects for a small region around Tharsis Tholus (42 images cartographically controlled as described in Robbins *et al.* (in rev.)). The top row shows results without flat-fielding, the bottom row shows results with flat-fielding. Panels A and D are "raw" mosaics with DN values based on standard calibration. Panels B and E show equalization results, and panels C and F show an average of those equalized results.

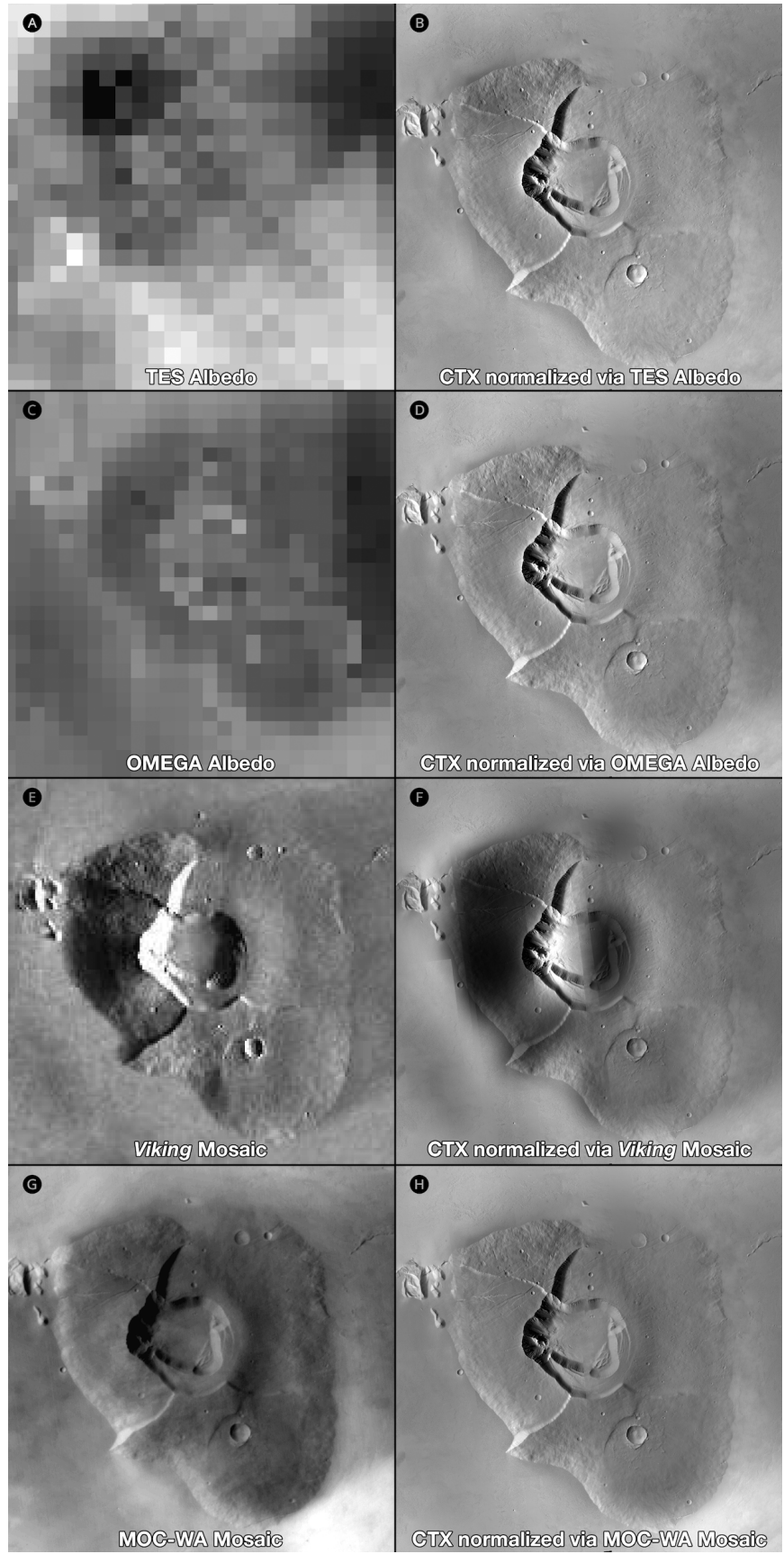

Figure 3.  Photometric correction of Fig. 2 using different references.  The reference for the region is in the left column, while the result is in the right column.  (A) is TES, (C) is OMEGA, (E) is *Viking*, and (G) is MOC-WA.  The small offsets in the left column are because none of the data are cartographically controlled, except *Viking*, with which the CTX aligns correctly in the right column.  Photometrically, *Viking* does the worst due to the different time-of-day.  While TES, OMEGA, and MOC-WA all do reasonably well in *this* area, problems are more obvious in broader views (not shown).

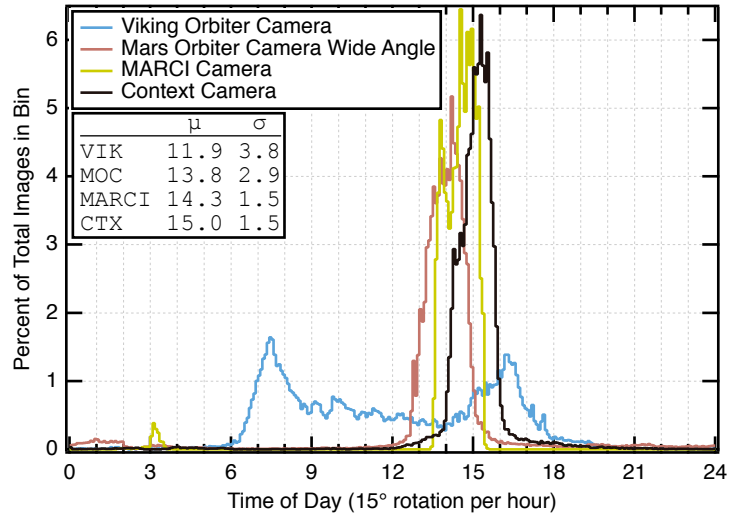

Figure 4.  Histogram of time-of-day images were acquired by spacecraft for possible reference relative to CTX.  (Note: *Viking* metadata data do not exist in PDS, so the images were obtained and processed to extract these data; however, only ~⅔ of the images could be successfully processed and only those are included here.)

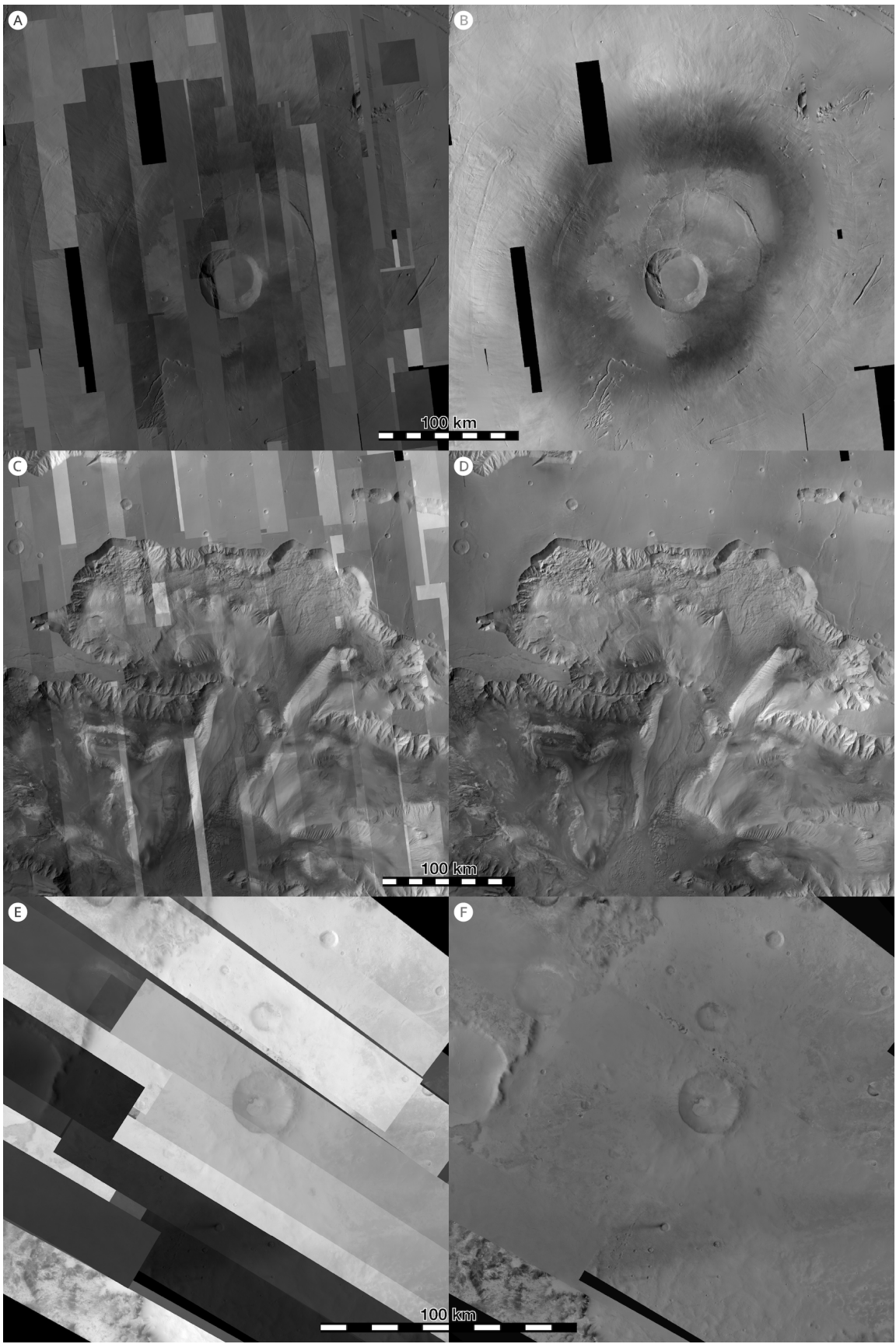

Figure 5.  Examples of our process applied to three different locations on Mars.  All images are stretched to 0.01%–99.99% of the dynamic range in the mosaic (which extend beyond what is shown).  Left column shows a "raw" mosaic (no corrections applied other than flat-fielding), and right column shows the empirical photometric control.  (A)–(B) Mosaic of 214 images around Pavonis Mons, and (C)–(D) mosaic of 162 images centered on Ophir Chasma, both cartographically controlled in Robbins *et al.* (2019), and photometrically controlled using the public 256ppd MOC-WA mosaic.  (E)–(F) Mosaic of 62 images in an area near the south pole (centered –70°N, 315°E), cartographically controlled in Robbins *et al.* (in rev.), and photometrically controlled using a 9ppd MOC-WA mosaic from Robbins (in rev.).